**Title**: Classifying ecosystem disservices and comparing their effects with ecosystem services in Beijing, China

**Authors:** Shuyao Wu[1*], Jiao Huang[2], Shuangcheng Li[2]

**Authors affiliations:**

[1] Environment Research Center, Duke Kunshan University

[2] Key Laboratory for Earth Surface Processes of the Ministry of Education Center of Land Science, College of Urban and Environmental Sciences, Peking University, Beijing, China

**\* Corresponding to:** E-mail: shuyao.wu@dukekunshan.edu.cn

Abstract:

Ecosystem disservices (EDS) is an important form of social-ecological interactions and can strongly influence people's perception of nature. However, compare to ecosystem services (ES) studies, current studies on EDS are still very limited especially from the perspective of classification and valuation. Since urban environment is a major venue of human-nature interaction, we used Beijing, the capital of China, as a case study area to value three common urban EDS (decrease in water quantity, increase in medical costs and infrastructure damage) and compare the results with the values of six ecosystem services (food and raw material production, climate regulation, environmental quality regulation, soil retention and ecotourism) to better understand the effects of EDS. The valuation results suggested that EDS and ES in Beijing were 203.4 billion and 9.12 billion RMB/year in 2018, respectively. The finding suggested that although EDS caused considerable financial loss, the potential economic gain from ecosystem services still greatly outweigh the loss and therefore supported the current urban greening expansion policy in Beijing. Our study attempted to promote the bridging of ecosystem services and disservices researches. We call for more equal consideration of both ES and EDS in ecosystem valuation studies in the future for more compressive and sustainable development plans.

**Introduction**

The research on ecosystem services (ES), which refers to the benefits people obtain from nature, has undergone tremendous progress in the past two decades (Costanza et al., 2017). In 2018 alone, there are over 4,800 ES related studies have been published in journals worldwide. On the other hand, the study on ecosystem disservices (EDS), which is defined as the negative effects of nature on human wellbeing, draws much less research attention by comparison (Shackleton et al., 2016). The first EDS related study was available in 2006 and only 46 literature published on this topic in 2018 (Blanco et al., 2019). Although the concept of EDS has been applied in the systems of agricultural, forestry and aquatic, the study area for most EDS literature is generally in cities, western Europe or the USA cities to be specific (Gomez-Gaggethun and Barton, 2013; Dohren and Haase, 2015).

But recently, more and more studies start to advocate the importance of expanding the research on EDS over the world. Blanco et al. (2019) proposed two very practical reasons for studying EDS. Firstly, since ES and EDS are distinct from and complementary to each other, studying EDS will improve our understanding of important social-ecological interactions, which would help people achieve sustainability. Secondly, since studies showed that stakeholders' actions could be more influenced by EDS than by ES, targeting EDS reduction might be a more effective way to promote nature-friendly and sustainable societies (Blanco et al., 2019).

In addition to the limited number of EDS studies and skewed study area coverage, current existing EDS research mainly focuses on raising the attention of EDS, defining and describing various EDS and quantifying their effects (Dunn, 2010; Lyytimaki, 2014). For instance, Wang et al. (2015) evaluated the relationship between plant diversity and ecosystem services and disservices provision ability in Beijing. Vaz et al. (2017) used the example of plant invasion to clarify the difference between ecosystem services and disservices and describe a framework that can integrate EDS into human wellbeing study. Speak et al. (2018) constructed a compound indicator system to compare the ES and EDS provided by urban trees and acquired the net benefits of urban ecosystems. Juanita et al. (2019) utilized expert knowledge to assess the impacts of land cover changes on ecosystem services and disservices provision in a Colombian city. Although these studies clarified the importance of EDS and improved our understanding of their effects, the exploration of the formation and valuation of ecosystem disservices just began (Ninan and Kontoleon, 2016; Shackleton et al., 2016).

Furthermore, although whether the effects of ecosystems should be monetized remains controversial (Read and Cato, 2014), the valuation of ecosystem disservices can play an important role in the areas such as policy decision-making, environmental cost-benefit analysis and environmental impact assessment (Kallis et al., 2013; Gunton et al., 2017). Therefore, Ninan and Inoue (2013) and Schaubroeck (2017) both argued that there is a need for equal consideration of ecosystem disservices and services when valuing nature to fully understand the overall effects of ecosystems to wellbeing. However, apart from Ninan and Kontoleon (2016)'s attempt to value the two forest ecosystem disservices in a protected area in India, a very limited amount of study assessed the net values of ecosystem effects.

Here, we aim to narrow these knowledge gaps by proposing a framework that may improve our understanding of the formation and types of ecosystem disservices. Then, we used Beijing, which is one of the largest metropolises in China, as a study area and attempt to value both important ecosystem services and disservices there. Specifically, we firstly applied the concept of cascade to delineate how ecosystem leads to human value loss to better understand the relationship between ecosystem structure, functions, disservices and human wellbeing, Secondly, we proposed two classification systems that based on the effect directness and functions of disservices to better apprehend the characteristics of ecosystem disservices. Finally, we estimated the values of three important ecosystem disservices in Beijing, China and compared them to the ecosystem services values for better understanding the net effects of urban ecosystems.

**Ecosystem disservices cascade and classification**

*Ecosystem disservices cascade*

Ecosystem disservices can be found in various forms. For example, Döhren and Haase (2015) summarized at least 14 urban ecosystem disservices found in literature, such as plants caused allergies, decrease in air quality, block of views, maintenance costs, infrastructure damage, introduction of invasive species, displacement of endemic species, etc. Similar to ecosystem services, the ecosystem structure and processes should also be the source of the various ecosystem disservices (Shackleton et al., 2016; Campagne et al., 2018). Therefore, accurate assessment and valuation of these various EDS rely on a clear understanding of how ecosystem structure and process negatively affect human wellbeing.

One way to delineate the mechanisms that underlie the ecosystem disservice formation is using the cascade diagram. Similar to the ES cascade proposed by Haines-Young and Potschin (2010), the EDS cascade also delineates how disservices are derived from ecosystem structure, processes and functions (Shackleton et al., 2016). Some of the functions that are beneficial to humans and then become ecosystem services. But some other functions or sometimes the same ones can also be harmful to humans, which ensues ecosystem disservices. In addition, some of the functions not only are unwelcomed by people but also negatively affect the biophysical structure of the ecosystem. For instance, there is mounting evidence showing that species invasion can reduce biodiversity, lower soil quality, increase disturbance frequencies and is usually regarded as a disservice by people (Rahlao et al., 2009; Duchicela et al., 2012; Hansen et al., 2018). However, it should be noted that the cascade effects of ecosystem disservices are dynamic and can change with different stakeholders and temporal/spatial scales (Shackleton et al, 2016; Campagne et al., 2018). Therefore, the cascade of any specific disservice only describes the formation of that disservice in a specific context of environment.

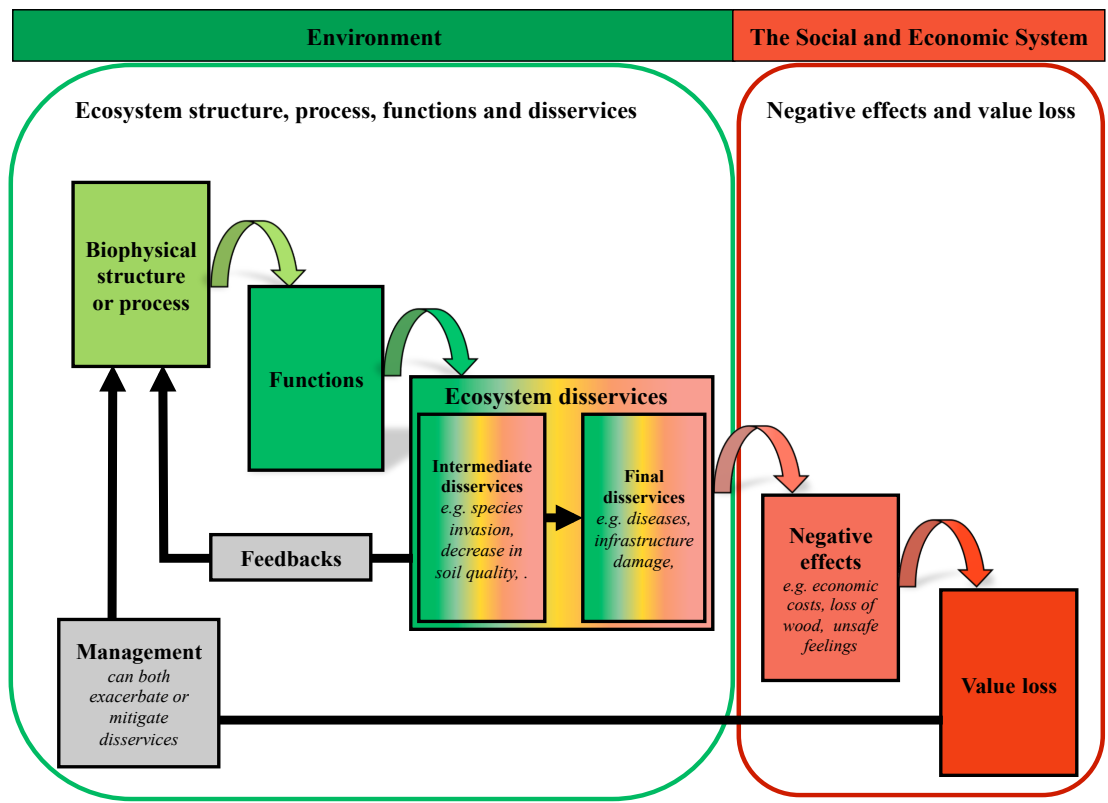

Fig. 1 The relationship between ecosystem structure, functions, disservices, negative effects and value loss.

*Intermediate and final ecosystem disservices*

For ecosystem services, Fisher et al. (2009) advocated the importance to differentiate intermediate and final services according to their degree of connection to human welfare in order to avoid double-counting in valuation. Similarly, the valuation of EDS also requires a distinction between intermediate and final EDS. Here, for the purpose of valuation, we suggest that to differentiate intermediate and final EDS based on whether they cause direct negative effects on human wellbeing (Table 1).

Final ecosystem disservices can be defined as the disservices that cause direct negative effects on human wellbeing. The direct negative effects refer to things such as financial costs, loss of goods, loss of revenue, unpleasant feelings, etc. Accordingly, some examples of final EDS can include a decrease in environment quantity, diseases, injuries, spending on infrastructure damage repair, unsafe feelings, etc. Although the maintenance costs for ecosystems are also a direct financial cost, they are not caused by any disservices from ecosystems. These final EDS are often excludable and/or rival, which makes their valuation results more reliable (Boyd and Banzhaf, 2007; Costanza, 2008; Fisher et al., 2009). By delineating a simplified scheme of the pathways of four common EDS in cities (i.e. infrastructure damage, decrease in water quantity and diseases or injuries), we can see how these EDS negatively affect people through adding direct financial costs (Fig. 2).

On the other hand, although intermediate EDS can negatively impact human welfare, they achieve these effects indirectly through increasing the delivery of final EDS and/or decreasing the provision of ecosystem services. Taking the introduction of invasive species as an example, the introduction of invasive species is a process that can decrease ecosystem productivity (Litton et al., 2006; Matthews and Spyreas, 2010; Hansen et al., 2018), water availability (Cordell and Sandquist, 2008), biodiversity (Healey and Gara, 2003; Tognetti et al., 2010; Tognetti and Chaneton, 2012; Herrera et al., 2016), which are the direct negative effects on human welfare (Wallace, 2007) (Fig. 2). Unless direct financial costs were spent on invasive species treatment (Olson, 2006; Pimentel et al., 2006), they should be regarded as a type of intermediate EDS. Some other examples of intermediate EDS include displacement of native species, decrease in soil/air/water quality, etc. Compare to final EDS, the values of these intermediate EDS are not independent but embed in the values of their corresponding final EDS and/or ES.

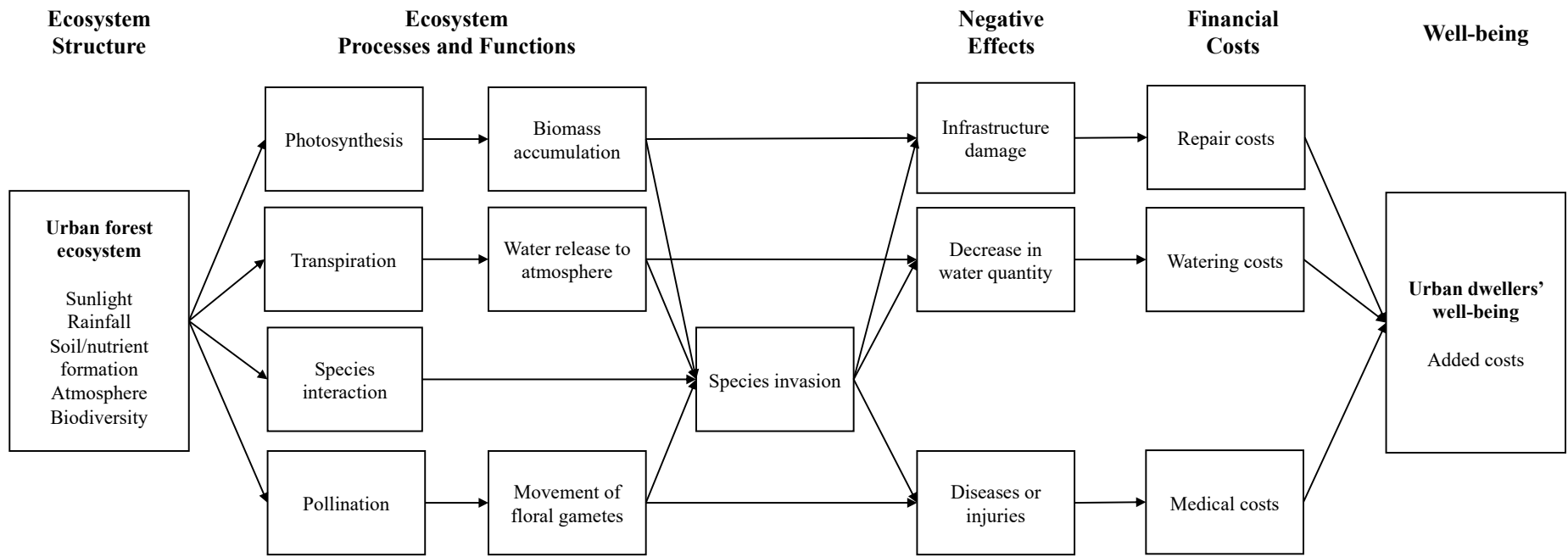

Fig. 2 Simplified scheme of the ecosystem pathways of three common ecosystem disservices found in the urban environment (infrastructure damage, decrease in water quantity and diseases or injuries).

*Ecosystem disservices functional classification*

The classification of ecosystem services helps people to better comprehend the complexity of ecosystem effects (de Groot et al., 2002; Costanza, 2008). The classification proposed by Millennium Ecosystem Assessment is one of the most commonly used classification systems of ES based on the functions due to its clarity and effectiveness (MA, 2005; Costanza et al., 2017). For ecosystem disservices, Lyytimäki (2014) attempted to classify dozens of EDS reported in newspapers into six groups, namely weather-related events, fears and risks, aesthetic issues, inhibition of activities and ecosystem functions causing harm. Shackleton et al. (2016) also divided ecosystem disservices into six categories based on their ecosystem origin and the dimension of human wellbeing affected. Moreover, Vas et al. (2017) suggested that the EDS could be categorized as five types, namely health, material, security and safety, cultural and aesthetic and leisure and recreation EDS.

Although these classifications of EDS have their merits, they are hard to compare with other ES study results. We believe a similar function-based EDS classification can help better incorporate the studies of ES and EDS. Therefore, we propose a functional classification system of EDS similar to the widely used ES classification of MA (Table 2). In this classification of EDS, provisioning EDS applies when the products people need but lost due to ecosystem functions and/or processes. Regulating EDS is defined as the harm or costs that people obtained from the regulation of ecosystem processes. Lastly, cultural EDS refers to nonmaterial harm or cost people obtained from ecosystems. Supporting EDS was not included since they are also the foundation for the provision of ES and very hard to identify independently. The examples of each type of EDS and their related ecosystem functions suggested indicators and the possible valuation approaches were also given (Table 2). In order to obtain more valid valuation results, we only provided examples of the final EDS valuation.

Table 1. Definitions and examples of intermediate and final ecosystem disservices

| Ecosystem Disservices Categories | Definitions | Examples | References |
| --- | --- | --- | --- |
| **Final Ecosystem Disservices** | The ecosystem disservices that cause direct negative effects on human wellbeing | Decrease in water quantity/diseases or injuries/infrastructure damage/unpleasant feelings | Geron et al., 1994; D'amato, 2000; Lyytimäki et al., 2008; Chaparro and Terradas, 2009; Lyytimäki and Sipilä, 2009; Dobbs et al., 2011; Escobedo et al., 2011; Pataki et al., 2011; Douglas, 2012; Nowak, 2012; Roy et al., 2012; Gómez-Baggethun and Barton, 2013; Kabisch and Haase, 2013; Seamans, 2013; von Dohren and Haase, 2015 |
| **Intermediate Ecosystem Disservices** | The ecosystem disservices that cause indirect negative effects on human wellbeing through increasing the delivery of final ecosystem disservices and/or decreasing the provision of ecosystem services | Introduction of invasive species/displacement of endemic species/decrease in soil nutrients/decrease in air quality/decrease in water quality | Adam and Boyle, 1982; Rothstein and Spaulding, 2010; Escobedo et al., 2011; Roy et al., 2012; von Dohren and Haase, 2015 |

Table 2. Types, definitions, examples, indicators and possible valuation approach for some final ecosystem disservices that caused direct negative effects on human wellbeing in an urban environment.

| Ecosystem Disservices Types | Definitions | Examples | Possible Indicators | Possible Valuation Approaches | References |
|---|---|---|---|---|---|
| **Provisioning disservices** | The products people need but lost due to ecosystem processes and functions | Decrease in water quantity by plants and wildlife | Amount of water that ecosystem needed from people, e.g. watering | *Added costs*: costs of the amount of water that consumed by plants and wildlife from people | Escobedo et al., 2011; Pataki et al., 2011; Roy et al., 2012; Gómez-Baggethun and Barton, 2013; Seamans, 2013; Dohren and Haase, 2015 |
| | | Loss of wood due to decaying | Amount of wood decayed | *Loss of benefits*: wood prices of the lost amount of biomass | |
| **Regulating disservices** | The harm or cost people obtained from the regulation of ecosystem processes and | Diseases or injuries caused by plants and wildlife | Increases in the number of patients of plants and wildlife-related diseases (e.g. rabies, Lyme disease, allergic | *Added costs*: the amount of medical costs and amount of repair costs spent on the damage from plants or wildlife | Geron et al., 1994; D'amato, 2000; Lyytimäki et al., 2008; Chaparro and Terradas, 2009; Lyytimäki and Sipilä, 2009; Dobbs et al., 2011; Escobedo et al., 2011; Pataki et al., 2011; |
| | functions | | rhinitis, acute rhinosinusitis, asthma, atopic dermatitis, etc.) | | Arnold, 2012; Douglas, 2012; Nowak, 2012; Roy et al., 2012; Gómez-Baggethun and Barton, 2013; Kabisch and Haase, 2013; Seamans, 2013 |
| | | Infrastructure damage | Average size and degree of infrastructure damage due to plants or wildlife | | |
| **Cultural disservices** | The nonmaterial harm or cost people obtained from ecosystems | Unpleasant (e.g. unsafe, disgusting, anxious, ugly) feeling caused by plants or wildlife | Decreases in the number of visitors | *Loss of benefits*: loss of the revenue from potential visitors; decrease in willingness-to-pay under different scenarios; use choice experiments to quantify the loss of benefits | Bixler and Floyd, 1997; Bolund and Hunhammar, 1999; Savard et al., 2000; Lyytimäki et al., 2008; Roy et al., 2012; Gómez-Baggethun and Barton, 2013; Seamans, 2013; Berry et al., 2018; Soto et al., 2018 |

**Ecosystem disservices valuation**

Here we proposed two possible approaches to estimate the values of EDS, which are through the added costs or loss of benefits approach. We believe that these two approaches can reflect the changes in values of natural capital from EDS, which contain both use and non-use values (Gómez-Baggethun et al., 2010; Munasinghe, 2010; Gunton et al., 2017; Ninan and Kontoleon, 2019). It is worthy to clarify that the word value in "EDS values" only refers to a monetary unit without any ethical implication.

In practice, various valuation techniques, such as conventional and implicit markets, can be applied to quantify the added costs and loss of benefits of EDS. Actual behaviors-based valuation methods utilize changes, such as losses in production or revenue, increased financial spending and deterioration in health status, to reflect EDS values directly and indirectly (Berry et al., 2018). On the other hand, intended behavior-based valuation methods including contingent valuation and choice experiment methods can also be applied for EDS valuation (Venkatachalam, 2004; Rakotonarivo et al., 2016). For instance, the possible decreases in willingness-to-pay might be obtained to quantify the value of unpleasant feelings that caused by ecosystems. Last but not least, value transfer is also a possible option when the results from other valuation studies are applicable (Brouwer, 2000; Richardson et al., 2015).

**Ecosystem disservices valuation in Beijing**

*Methods*

Beijing is the capital of China and has a population of 21.54 million in 2018. It covers an area of 16,410 km$^2$ and contains over 895,000 ha of urban forestry and greening space in 2018 (Beijing Statistical Yearbook, 2019). The large population and green space create both high demand and supply of urban ecosystem services. However, at least three ecosystem disservices were also present in this city, which are infrastructure damage, diseases or injuries and decrease in water quantity.

The valuation approach for the three important urban ecosystem disservices was based on the added costs approach since these disservices directly led to real and measurable monetary expenditure. For the disservices that with available statistic data, such as the decrease in water quantity, we calculated its value based on the government-released statistics and public information. For the disservices without ready statistics (e.g. diseases or injuries caused by plants or wildlife), data from local studies were used for added cost estimation. For infrastructure damage, we used the value transfer method to obtain an approximate value since there is no local data available. The detailed calculation process for each disservice are as follows:

*Infrastructure damage*

The value of infrastructure damage measures the added costs people spent on repairing the damage due to the growth of plants on infrastructures, such as sidewalk, street pavement, curb, gutter and sewer (McPherson and Peper, 1996). Its valuation is based on the percentage of repair costs in terms of total maintenance costs. A value transfer technique is used to determine the percentage.

$$V_I = M \times P_T$$

where $V_I$ is the repair costs of the infrastructure damage caused by plants. $M$ refers to the maintenance costs spent on urban ecosystems in Beijing, which equals approximately 1.82 billion RMB/year in 2018. $P_T$ represents the percentage of repair costs in terms of the total maintenance costs based on a value transfer method. McPerson and Peper (1996) and McPerson (2000) assessed 33 USA and Canadian cities in total and found the damage costs equal to approximately 44% of the maintenance costs on average.

*Decrease in water quantity*

The value of the decrease in water quantity measures the added costs that people need to spend for compensating the deficit between natural water supply and water consumption by vegetation. Its valuation is based on the annual statistics of the amount of artificial watering for ecological and agricultural purposes and the corresponding local water prices.

$$V_W = A_E \times Pr_{WE} + A_A \times Pr_{WA}$$

where $V_W$ is the artificial watering costs for compensating the decrease in water quantity due to plant growth. $A_E$ and $A_A$ are the amounts of water consumed by plant growth and agricultural production in Beijing and equal to 1.34 billion m³ and 42 million m³ in 2018, respectively (Beijing Statistical Yearbook, 2019). $Pr_{WE}$ and $Pr_{WA}$ are the local prices of water for ecological and agricultural purposes and equal to 6 RMB/m³ and approximately 0.12 RMB/m³, respectively.

*Diseases or injuries caused by plants or wildlife*

The value of diseases or injuries measures the added costs that people spent on medication due to plants and wildlife. In Beijing, studies showed that asthma and allergic rhinitis (AR) are the two most common diseases induced by plants hence the focuses of the valuation here (Wang et al., 2016; Wang et al., 2017). The valuation of this disservice focus on quantifying the medical costs of plant-induced asthma and AR treatment and is based on the results of local studies.

$$V_D = \sum_{i=1}^{2} Pop \times \alpha_i \times \beta_i \times C_i$$

where $V_D$ is the medical costs of the diseases or injuries that caused by plants or wildlife. *Pop* is the population of Beijing in 2018, which is approximately 21.54 million people (Beijing Statistical Yearbook, 2019). $\alpha_i$ is the incidence rate of plant- or wildlife-related diseases or injuries $i$ in Beijing (%), $i$ refers to asthma and allergic rhinitis in this study. The incidence rate of asthma and allergic rhinitis in Beijing are about 0.81% and 1.29%, respectively based on the survey results of Wang et al. 2008. $\beta_i$ is the percent of patients of disease $i$ caused by plants in Beijing (%), which is about 61% (Li et al., 2015). $C_i$ represents the medical costs for each patient of disease $i$ (RMB/person). For asthma and allergic rhinitis, they are approximately 977.03 RMB/person and 629.68 RMB/person, respectively in China (Peng and Li, 2004; Chen and Li, 2014).

ES valuation

The values of ecosystem services in Beijing are derived from the 2018 Urban Modern Agricultural Ecosystem Services Value Annual Report (http://tjj.beijing.gov.cn/tjsj/tjgb/stgb/201905/t20190520_174010.html). It is a report released by the Beijing Statistic Bureau annually since 2006 and calculated the values of 12 ecosystem services provided by the forest, agricultural fields, grassland and wetland in Beijing. In order to avoid potential double-counting problem and make the comparison of ES and EDS values more valid, we only choose the values of a total of five provisioning, regulating and cultural services from the report. They are food and raw material production, climate regulation, environmental quality regulation (include air quality regulation, water quality regulation and noise reduction), soil retention and ecotourism. More information on the calculation methods of each ecosystem service can be found in the supplementary material.

*Results & Discussion*

According to the Urban Modern Agricultural Ecosystem Services Value Annual Report, the 2018 values of ecotourism, climate regulation, food and raw material production, environmental quality regulation and soil retention in Beijing are 85.5 billion, 76.3 billion, 29.7 billion, 11.6 billion, 296.8 million RMB/year, respectively (Fig.3a). The estimated value of ecosystem disservices, which are decrease in water quantity, infrastructure damage and increase in medical costs, are 8.1 billion, 798,9 million and 231.2 million RMB/year, respectively in 2018 (Fig. 3b). The total values of the ecosystem services and disservices are 203.4 billion and 9.12 billion RMB/year, respectively, which renders approximately 194.3 billion net value of the ecosystem effects in Beijing in 2018 (Fig. 4). Among the ES, ecotourism and climate regulation are the two services that have the highest monetary values, which account for 42% and 37% of the total ES value, respectively. Compare to the other two EDS, decrease in water quantity is the most important EDS in Beijing in terms of value, and accounts to about 89% of the total EDS value. It is reasonable given that the average 500mm annual precipitation in Beijing is certainly unable to keep up with the large plant water demand, especially under the pressure from both industrial and domestic water usage (Li et al., 2017).

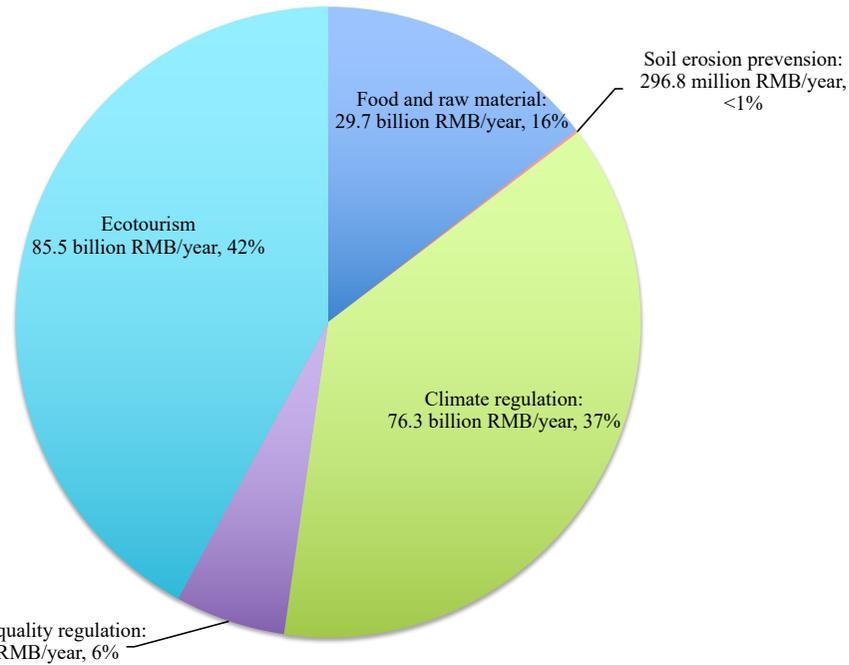 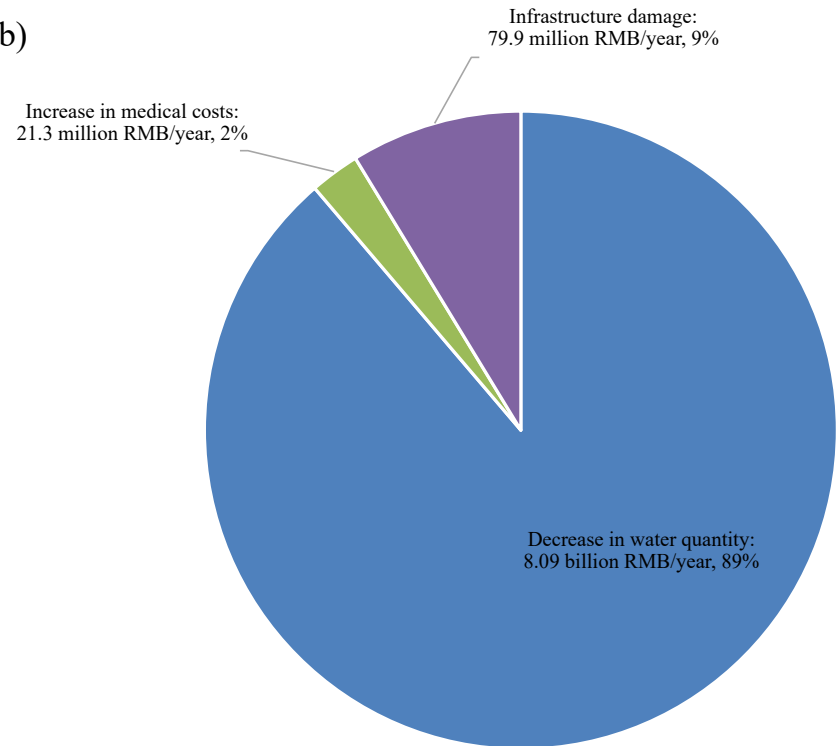

Fig. 3 Value composition of the five ecosystem services (a) (food and raw material production, climate regulation, environmental quality regulation, soil retention and ecotourism) and three disservices (b) (decrease in water quantity, increase in medical costs and infrastructure damage) in Beijing in 2018.

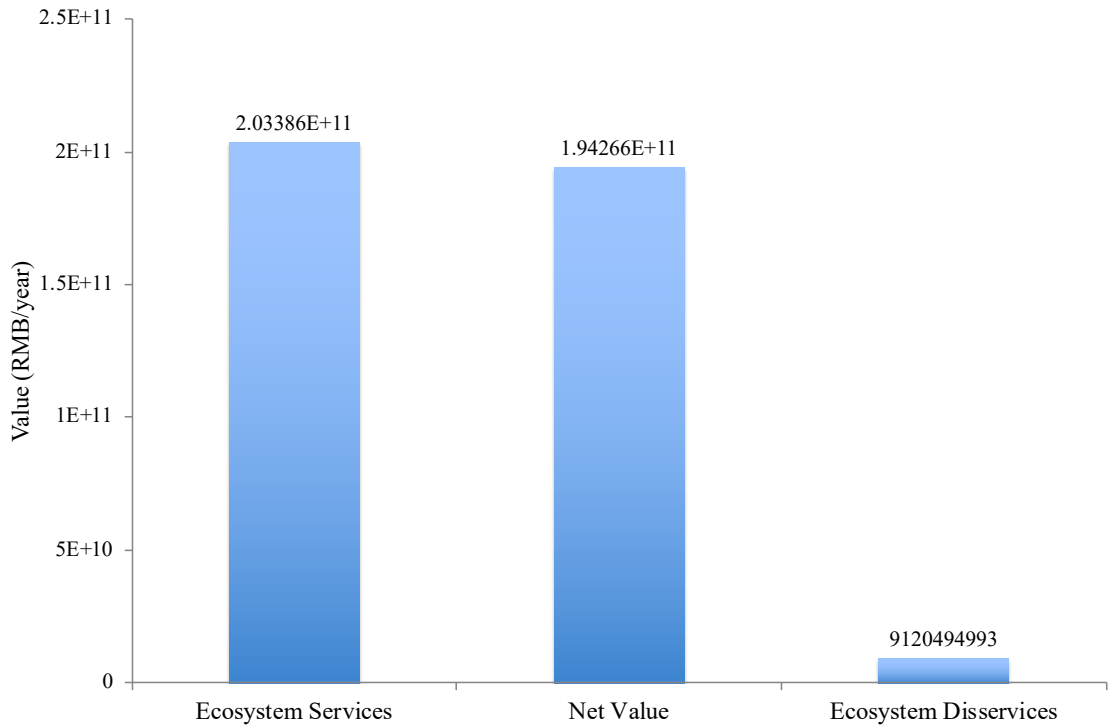

Fig. 4 Comparison of estimated values of five ecosystem services and three disservices and the net value of these urban ecosystem effects in Beijing, 2018.

For policy-making implications, the current urban ecosystem management policies of the Beijing municipal government mainly focus on increasing the forestry coverage. For example, the "New One Million Mu Urban Planting Plan of Beijing", which aims to increase the urban forest coverage by approximately 66,667 ha before 2023 (http://www.beijing.gov.cn/zfxxgk/110038/qtwj22/2019-04/08/content_35674cac1d7749b0897d0e9fb0721459.shtml). Our evaluation results of the net ecosystem effects support the decision since the net value of urban ecosystem are still very positive after considering both ES and EDS. The total value of EDS is 4.5% of the total ES values. However, the usage of water-consuming species should be reduced since they render a considerable amount of financial costs. Some of the most commonly planted water-consuming species in Beijing including *Salix alba, Acer truncatum* and *Malus micromalus* for examples (Wang, 2006). They may be switched to more water conservative native species, such as *Pinus tabuliformis*, *Platycladus orientalis and Cotinus coggygria* in order to minimize the effects of EDS (Wang, 2006; Che, 2008).

It is also important to keep in mind that the proposed EDS valuation methods also share similar caveats and limitations of ES valuation studies. For instance, the accuracy of the estimated EDS value heavily depends on the data quality and can change based on the applied prices and methods (Braat and de Groot, 2012). For example, the value of infrastructure damage of this study can change easily if different value transfer source is applied. Furthermore, the provision and value of EDS might also vary in time and space. These temporal and spatial heterogeneities were not captured in this study due to data limitations. Future studies should incorporate these heterogeneities while considering more types of EDS (e.g. unpleasant feelings, accidents, loss of carbon, etc.) and valuation methods (e.g. contingent valuation) to improve the completeness and accuracy of both ES and EDS valuation studies. In future studies, it is essential to address these issues and consider more ecosystem services (e.g. flood risk mitigation, soil quality regulation, noise attenuation, aesthetics, etc.) and disservices (e.g. unpleasant feelings, accidents, loss of carbon, etc.). It is also crucial to test more valuation methods, such as contingent valuation, choice experiment and hedonic pricing for their capability and validity on EDS valuation, in order to obtain more comprehensive and accurate assessments on the values of urban ecosystem effects.

**Conclusions**

To completely understand the effects of urban ecosystems, the effects of ecosystem disservices should be considered along with the ecosystem services and require more research attention. In this study, we tried to better understand its formation through the use of cascade flowchart and classification systems and compare their effects with ecosystem services. It is vitally important to differentiate final and intermediate ecosystem disservices for understanding the negative effects of the ecosystem on human well-being. The proposed functional classification of EDS (i.e. provisioning, regulating and cultural EDS) should also help better bridging EDS and ES studies. In addition, we used Beijing as a case study area to value the EDS caused by urban ecosystems and compare the findings with ES values. The results suggested that although EDS caused great financial loss the potential economic gain from ecosystem services still significantly outweigh the loss. Our study only sheds light on valuating the net effects of urban ecosystems. In the future, we believe that EDS valuation should be at least equally considered in ecosystem valuation studies to create more comprehensive and sustainable development policies, land use proposals and management plans.

## Supplementary Material

A series of production-based and cost-based valuation methods were applied to estimate the 2018 values of five major ecosystem services in Beijing, including market prices (for food and raw material production and ecotourism), replacement costs (for climate regulation and environmental quality regulation) and avoided damage costs (for soil retention services). The data required for the valuation were obtained by the statistics data from various government agencies, such as the Beijing Gardening and Greening Bureau, Beijing Water Authority, Beijing Municipal Bureau of Agriculture and Rural Affairs and Beijing Municipal Bureau of Statistics. The authors talked to the Beijing Municipal Bureau of Statistics and obtained the information on the detailed valuation process for each service as follows:

Food and raw material production:

$$V_F = \sum_{i=1}^{4} Pro_i \times Pr_i$$

where $V_F$ is the estimated service value of food and raw material production (RMB/year). $Pro_i$ is the annual production of food or raw material $i$ (t/year), $i$ refers to either agricultural products, wood products, husbandry products or fishery products. $Pr_i$ is the price of food or raw material $i$ (RMB/t).

Climate regulation:

$$V_{Cli} = V_T + V_H + V_{CO2} + V_{O2}$$
$$V_T = A_W \times ET_{avg} \times Va \div Ef \times Pr_E$$
$$V_H = A_W \times ET_{avg} \times X \times Pr_E$$
$$V_{CO2} = \sum_{i=1}^{3} Carbon_{iC} \times Pr_{iC}$$
$$V_{O2} = \sum_{i=1}^{4} Oxygen_{iO} \times Pr_{iO}$$

where $V_{Cli}$ refers to the value of climate regulation (RMB/year), which is the sum of value of temperature regulation ($V_T$, RMB/year), humidity regulation ($V_H$, RMB/year), sequestrated carbon ($V_{CO2}$, RMB/year) and released oxygen ($V_{O2}$, RMB/year). $A_W$ is the sum of wetland area of Beijing (ha), $ET_{avg}$ represents the long-term annual average evaporation amount of wetland in Beijing (mm). $Va$ is the heat of vaporization of water (kJ/kg). $Ef$ refers to the average energy efficiency of air conditioners. $X$ refers to the amount of electricity needed to evaporate 1 m³ water. $Pr_E$ is the price of electricity in Beijing (RMB/kWh). $Carbon_{iC}$ is the amount of carbon sequestrated in forest, grassland and agricultural fields. $Oxygen_{iO}$ is the amount of oxygen released by wetland, forest, grassland and agricultural fields. $Pr_{iC}$ and $Pr_{iO}$ refer to the costs for artificial carbon sequestration and oxygen production, respectively.

Environmental quality regulation (air quality regulation):
$$V_A = V_R + V_I$$
$$V_R = R \times Pr_R$$
$$V_I = I \times Pr_I$$

where $V_A$ is to the value of air quality regulation (RMB/year), which is the sum of value of pollutant reduction ($V_R$, RMB/year) and air quality improvement ($V_I$, RMB/year). $R$ refers to the estimated total amount of reduced pollutants by forest, agricultural fields and grassland (t/year) and $Pr_R$ is the cost needed for the same amount of reduction through artificial methods (RMB/t). $I$ is the estimated total amount of anion released by forest and wetland (t/year) and $Pr_I$ is the cost needed for the same amount of production through artificial methods (RMB/t).

Environmental quality regulation (water quality regulation):
$$V_{WQ} = (W_F + W_W) \times Pr_{WQ}$$

where $V_{WQ}$ refers to the value of water quality regulation (RMB/year). $W_F$ and $W_W$ are the water purification ability of forest and wetland, respectively (t/year). $Pr_{WQ}$ is the cost needed for the same amount of purification through artificial methods (RMB/t).

Environmental quality regulation (noise reduction):
$$V_N = (N_F \div N_R) \times Pr_N$$

where $V_N$ represents the value of noise reduction (RMB/year). $N_F$ and $N_R$ are the noise reduction ability of forest and soundproof windows (dB/year), respectively. $Pr_N$ is cost of soundproof windows (RMB/dB).

Soil retention:
$$V_S = V_{SN} + V_{SC}$$
$$V_{SN} = R_S \times N_{SN} \times Pr_{SN}$$
$$V_{SC} = R_S \div \rho_S \times P_{SC} \times Pr_{SC}$$

where $V_S$ is to the value of soil retention (RMB/year), which is the sum of value of soil nutrient retention ($V_{SN}$, RMB/year) and avoided cleaning cost ($V_{SC}$, RMB/year). $R_S$ is the amount of soil retained by vegetation (t/year). $N_{SN}$ represents the average nutrient content of soil. $\rho_S$ is the bulk density of soil (t/m³). $P_{SC}$ is the percentage of soil that would be retained in water reservoirs or channels without vegetation (%). $Pr_{SN}$ and $Pr_{SC}$ are the prices of soil nutrient fertilizers and soil cleaning costs (RMB/t).

Ecotourism:
$$V_{Eco} = V_{Rec} + V_{Edu}$$
$$V_{Rec} = \sum_{i=1}^{N} Visitor_{iV} \times Pr_{iV}$$
$$V_{Edu} = \sum_{i=1}^{N} A_{iEdu} \times X_{iEdu} \times In_{Edu}$$

where $V_{Eco}$ is to the value of ecotourism (RMB/year), which is the sum of value of recreation ($V_{Rec}$, RMB/year) and education ($V_{Edu}$, RMB/year). $Visitor_{iV}$ stands for the annual number of visitors of one park or nature reserve $iV$. $Pr_{iV}$ refers to the entrance fee per person of the park or nature reserve $iV$ (RMB/person). $A_{iEdu}$ is the area of one nature reserve or wetland $iEdu$ (ha). $X_{iEdu}$ is an adjustment coefficient based on the characters of nature reserve or wetland $iEdu$. $In_{Edu}$ is the annual education income per hectare of nature reserves and wetlands (RMB/ha/year).